\documentclass[10pt,conference]{IEEEtran}
\IEEEoverridecommandlockouts

\usepackage{cite}
\usepackage{amsmath,amssymb,amsfonts}
\usepackage{amsthm}
\usepackage{algorithmic}
\usepackage{algorithm}
\usepackage{graphicx}
\usepackage{textcomp}
\usepackage{xcolor}
\usepackage{balance}
\usepackage[a4paper,left=1.6cm,right=1.6cm,top=2cm,bottom=4.4cm]{geometry}

\def\BibTeX{{\rm B\kern-.05em{\sc i\kern-.025em b}\kern-.08em
    T\kern-.1667em\lower.7ex\hbox{E}\kern-.125emX}}
\begin{document}

\title{Source PAC Coding for Low-latency Secret Key Generation in Short Blocklength Regime}


\author{\IEEEauthorblockN{Lulu Song\IEEEauthorrefmark{1},
Di Zhang\IEEEauthorrefmark{2}, Tingting Zhang\IEEEauthorrefmark{3} }
\IEEEauthorblockA{\IEEEauthorrefmark{1}School of Electrical and Information Engineering, Zhengzhou University, Zhengzhou 450001, China}
\IEEEauthorblockA{\IEEEauthorrefmark{2}School of Intelligent Systems Engineering, Sun Yat-sen University, Shenzhen 518107, China}
\IEEEauthorblockA{\IEEEauthorrefmark{3}Key Laboratory of Maritime Intelligent Cyberspace Technology of Ministry of Education, Hohai University,\\ Changzhou 213200, China}
\IEEEauthorblockA{Corresponding author: Di Zhang}
}

\maketitle

\begin{abstract}
Source polar coding is a potential solution for short blocklength-based low-latency key generation with limited sources, which is a critical aspect of six generation (6G) Internet of things. However, existing source coding schemes still suffer from significant degradation in key generation rate and reconciliation reliability in short blocklength regime. To address this issue, we introduce a multilevel source polarization-adjusted convolutional (PAC) coding framework. Furthermore, we propose a novel code construction algorithm that jointly leverages polarization effects and the maximum likelihood (ML) decoding error coefficient. Simulations demonstrate that the multilevel source PAC scheme with the proposed code construction achieves superior key generation rate under key disagreement constraints compared to conventional and multilevel source polar coding methods even in short blocklength regimes.
\end{abstract}


\section{Introduction}
The proliferation of resource-constrained devices and real-time applications has intensified demand for low-latency key generation schemes operating under limited source observations \cite{one-shot}, which play a pivotal role in six generation (6G) Internet of Things (IoT), enabling real-time, trustworthy data exchange essential for mission-critical applications such as autonomous systems, remote surgery, and smart infrastructure \cite{6G}. As a promising solution for scenarios like wireless physical-layer security and quantum key distribution, source coding with side information extracts secret keys from correlated source observations \cite{lattices}. Typically, the number of source observation pairs and the blocklength in source coding are assumed to be asymptotically large. However, limited numbers of source observations necessitate short blocklength source coding schemes, which is hard to simultaneously achieve a satisfactory key generation rate and disagreement rate\cite{commcontrainedkey}.

Source polar coding, known to asymptotically achieve secret key capacity for discrete memoryless sources, has been investigated in key reconciliation \cite{chousourcepolar}. Prior study in \cite{shortpolar} indicates its competitive performance even at short blocklength regimes. Nevertheless, its reconciliation efficiency remains suboptimal due to incomplete polarization in these short blocklength regimes and inferior distance properties. Introduced by Arıkan, polarization-adjusted convolutional (PAC) codes demonstrate enhanced short blocklength performance compared to standalone polar codes \cite{rowmerged}. By incorporating a rate-1 convolutional code as the outer code, PAC codes enable a better utilization of the incompletely polarized bit-channel and improve the distance spectra.

As the key determinant of the PAC codes performance, a good code construction is required to select the frozen and unfrozen bits appropriately. Polar construction \cite{polarrule1} maximizes polarization utilization but often yields poor distance spectra. Conversely, Reed-Muller (RM) construction \cite{rowmerged} optimizes maximum likelihood (ML) performance but restricts code rates and under-utilizes the polarization. Hybrid approaches like RM-polar \cite{rmpolar} and the construction in \cite{Chiupolarml} attempt to balance both aspects but still suffer inherent deficiencies: RM-polar loses its hybrid advantage and degenerates to classical RM at rational rates $R = a/b$ where $b$ is a power of two; the latter adopts the weight enumerating function averaged over the ensemble of random PAC codes, which is inaccurate and is followed by a complex and suboptimal optimization of the convolutional code. Consequently, extending PAC principles to source coding and developing effective constructions remain open challenges.

In this work, a low-latency key generation scheme under limited numbers of continuous source observations is proposed. A novel multilevel source PAC code for key reconciliation is designed to mitigate the degradation in key rate and key disagreement in short blocklength regime. With the source coding problem with side information treated as a channel coding problem, the virtual channels are transformed to binary-input additive white Gaussian noise (BI-AWGN) channels with equivalent capacities. Then the minimum Hamming distance and its multiplicity determining the ML decoding performance are investigated. A code construction algorithm is proposed to utilize both the polarization effect and the ML decoding performance through Gaussian approximation (GA) and efficient semi-closed-form enumeration. Simulation results demonstrate that this method can significantly improve the key generation rate compared to conventional and multilevel source polar codes.

\section{Problem Description}
Given a scenario in which Alice and Bob aim at negotiating a key $S$ from correlated source observations to prevent Eve from eavesdropping. Assume that Alice and Bob observe $N$ i.i.d. correlated variable pairs $(h_A, h_B)$, where
\begin{equation}
h_A = h + n_A, h_B = h + n_B,
\end{equation}
and $h \sim \mathcal{CN}(0, \sigma_h^2)$. $n_A, n_B \sim \mathcal{CN}(0, \sigma_n^2)$ are additive noise independent of $h$. Alice extracts a key $S_{1:K}$ and a public message $F$ from $h_A^{(1:N)}$ and sends $F$ to Bob via a public noiseless channel. Bob estimates the key $\hat{S}_{1:K}$ according to $h_B^{(1:N)}$ and the received $F$. Assume that Eve has only access to $F$ and the statistics of $h_A$ and $h_B$. 

The key metrics of key generation problems include the generation rate, reliability and information secrecy of the generated key. The generation rate is defined as the ratio of the key length and the number of source variables:
\begin{equation}
    R_{K} = K/N.
\end{equation}
The reliability is measured by the key disagreement rate (\text{KDR}):
\begin{equation}
    \text{KDR} = \Pr(S \neq \hat{S}).
\end{equation}
Considering the bit error rate in one-time pad encryption, the reliability can also be measured by the bit disagreement rate (\text{BDR}):
\begin{equation}
    \text{BDR} = \frac{\sum_{k=1}^{K} \left| S_k - \hat{S}_k \right|}{K}.
\end{equation}
The secrecy is measured by the total variance distance between the actual and ideal joint distributions of $S \times F$, which is a strong secrecy metric in the sense of information theory:
\begin{equation}
\begin{split}
    D_{SF} & = d(P_{SF}, Q_{\mathcal{K}}^{\text{unif}} \times P_F)\\
    & \triangleq \frac{1}{2} \sum_{s \times f \in \mathcal{S\times F}} |P_{SF}(sf) - Q_{\mathcal{K}}^{\text{unif}} \times P_F(sf)|\\
    &= \sup_{\mathcal{A} \subseteq \mathcal{S\times F}} |P_{SF}[\mathcal{A}] - Q_{\mathcal{K}}^{\text{unif}} \times P_F[\mathcal{A}]|,
\end{split}
\end{equation}
where $P_{SF}$ and $P_{F}$ denote the distributions of $S \times F$ and $F$, respectively. $Q_{\mathcal{K}}^{\text{unif}}$ characterizes the uniform distribution over the underlying space $\mathcal{K} \triangleq \{0,1\}^K$.

Asymptotically, when $N \rightarrow \infty$, the maximum achievable rate, i.e. the key capacity with $\epsilon \rightarrow 0$, $\delta \rightarrow 0$ is
\begin{equation}
    C_K = I(h_A,h_B) = \log(1+\frac{\sigma_h^{4}}{2\sigma_h^{2}\sigma_n^{2}+\sigma_n^{4}}).
\end{equation}
However, observing large numbers of source variable pairs takes a long time before the required key can be generated and incurs excessive latency. This contradicts the low-latency constraints of key generation in time-sensitive applications. Therefore, we consider the problem under the assumption of limited source observations. With a limited and particularly short length $N$, the asymptotic bound is inaccurate. By introducing a binary hypothesis test that distinguishes between two distributions, a tight upper bound that depends on $N$ can be derived as \cite{shortpolar}
\begin{equation}
\begin{split}
    R_{\text{K}} &\lessapprox I(h_A; h_B) + 2\sqrt{\frac{V}{N}}Q^{-1}(1 - \epsilon - \delta - \tau) \\
    &+ \frac{2}{N} \left( \log \frac{\tau + \delta}{\tau} + \frac{1}{2} \log N \right),
\end{split}
\end{equation}
where $\epsilon > 0$, $\delta > 0$, $0 \leq \epsilon + \delta < 1$ and $0 < \tau < 1 - \epsilon - \delta$. $Q\left( x \right) {=} \int_x^\infty {\frac{1}{{\sqrt {2\pi } }}} {e^{ - {t^2}/2}}dt$ is the Gaussian $Q$ function and $V = \frac{\sigma_h^4}{(\sigma_h^2 + \sigma_n^2)^2} \log^2e$.

This work aims to develop a low-lantency key generation scheme under short source length $N$, and improve the key generation rate $R_K$ toward the theoretical upper bound under the reliability constraint $\text{KDR} \leq \epsilon_{\text{KDR}}$ (or $\text{BDR} \leq \epsilon_{\text{BDR}}$) with the help of the proposed short blocklength source coding architecture. The details of the scheme are elaborated in Section III and Section IV.

\section{Key Generation scheme}
\subsection{Alice's Side}
\subsubsection{Quantization and Labeling}
By splitting the real and imaginary components of $h_A^{(1:N)}$, Alice can obtain a real-valued sequence $X_{1:2N}$ where $X \sim \mathcal{N}(0, \sigma_h^2+\sigma_n^2)$. For further process, quantization and labeling are needed to discrete the continuous space. Equiprobable quantization is adopted to ensure a uniform distribution over the key sample space $\mathcal{K}$. Assume $Q \in \mathbb{Z}^+$ quantization levels and $2^Q$ quantization intervals $R_t = [r_{t-1}, r_t], t=1,\dots,2^Q$. Then we have
\begin{equation}
\begin{split}
    \Pr(X \in R_t) &= \int_{r_{t-1}}^{r_{t}} \frac{1}{\sqrt{2\pi (\sigma_Z^2 + \sigma_W^2)}} e^{-\frac{x^2}{2(\sigma_Z^2 + \sigma_W^2)}} \, dx \\
    &= 1/2^Q, \quad \text{for } t = 1, \ldots, 2^Q
\end{split}
\end{equation}

After the determination of the quantization interval, labeling is needed to assign a label of $Q$ bits to each interval. The label for the $t$th interval is denoted as $label(R_t) = l_{1:Q}^t$. After equal probability quantization and Set Partitioning (SP) labeling, $x_{1:2N}$ is mapped to $l_{1:Q}^{(0:2N-1)}$. 

\subsubsection{Multilevel Source Polarization-adjusted Convolutional Coding}
To accommodate the continuous nature of the source and the short blocklength requirements, we introduce a multilevel source PAC coding architecture here. First, Alice extracts the level-$q$ sequence $l_q^{(0:2N-1)}$ from $l_{1:Q}^{(0:2N-1)}$. Then for every $q \in [1:Q]$, $l_q^{(0:2N-1)}$ is encoded as $u_{0:2N-1}^{(q)}$ through polar transformation
\begin{equation}
    u_{0:2N-1}^{(q)} = l_q^{(0:2N-1)} \mathbf{G}_p,
\end{equation}
where $\mathbf{G}_p = \mathbf{P}_{2N}\mathbf{G}^{\bigotimes \log{2N}}$ is the polarization generator matrix. $\mathbf{P} \in \mathbb{F}_2^{2N\times2N}$ is a permutation matrix with recursive expression $\mathbf{P}_{2N} = \mathbf{R}_{2N}(\mathbf{I}_2 \bigotimes \mathbf{P}_{2N/2})$, where $\mathbf{P}_2 = \mathbf{I}_2$, and $\mathbf{R}_{2N}$ represents the operation separating the odd-indexed and the even-indexed components. In addition to that, $\mathbf{G} = [\begin{smallmatrix}
    1 & 0\\
    1 & 1
\end{smallmatrix}]$ is the kernel matrix. Note that the permutation operation is essential here due to the introduction of convolutional coding.

Since the polarization takes place slowly in short blocklength regime, the $0-1$ assignment of frozen bits results in a waste of capacity and thus a loss in key generation rate. Therefore, we introduce a rate-one convolutional coding following the polar transformation as follows
\begin{equation}
    v_{0:2N-1}^{(q)} = u_{0:2N-1}^{(q)} \mathbf{C},
\end{equation}
where $\mathbf{C}$ is an upper triangular Toeplitz matrix generated by generator polynomial $c(D) = c_0 + c_1 D + \cdots + c_d D^d $ with constraint length $d+1$. Generally, we have $c_0 = c_d = 1$. In source PAC coding, the frozen bits are chosen from $v_{0:2N-1}^{(q)}$ rather than $u_{0:2N-1}^{(q)}$. As a result, $u_{0:2N-1}^{(q)}(\mathcal{I}_F^q)$ are dynamic frozen bits unknown before decoding and dependent on previous decoded bits, where $\mathcal{I}_F^q$ is the frozen set.

\subsubsection{Message Transmission and Key Extraction}
Next, the massage $f_{1:M} = [f_{1:M_1}^{(1)}, \dots, f_{1:M_q}^{(Q)}]$ to be transmitted to Bob and the key $s_{1:K} = [s_{1:K_1}^{(1)} \dots, s_{1:K_q}^{(Q)}]$ are extracted from $v_{0:2N-1}^{(1:Q)}$. For every $q \in \{1, \dots, Q\}$, we have
\begin{equation}
    \begin{split}
        f_{1:M_q}^{(q)} = v_{0:2N-1}^{(q)}(\mathcal{I}_{F}^{q}), \quad s_{1:K_q}^{(q)} = v_{0:2N-1}^{(q)}(\mathcal{I}_{S}^{q})
    \end{split}
\end{equation}
where $M_q$ and $K_q$ are the length of the massage and the key at level $q$. $\mathcal{I}_{F}^{q}$ and $\mathcal{I}_{S}^{q}$ represent the frozen set and the key set.

\subsection{Bob's Side}
\subsubsection{Multistage SCL Decoding}
Leveraging $f_{1:M}$ from Alice and its own observation $Y_{1:2N}$, Bob performs multistage SCL decoding to form an estimate $\hat{v}_{0:2N-1}^{(1:Q)}$ of $v_{0:2N-1}^{(1:Q)}$. At the $q$-th SCL decoding stage, Bob forms the estimate $\hat{v}_{0:2N-1}^{(q)}$ of $v_{0:2N-1}^{(q)}$ based on the received frozen bits $f_{1:M_q}^{(q)}$, local source observations $Y_{1:2N}$, and the estimate $\hat{l}_{1:q-1}^{(0:2N-1)}$ of $l_{1:q-1}^{(0:2N-1)}$ obtained from $q-1$ previous stages.

Different from channel PAC codes, in source PAC codes, the log-likelihood ratios (LLRs) computation flows recursively layer by layer from the source side to the encoded side. In the $q$-th decoding stage, the initial LLR at the source side is
\begin{equation}
\begin{split}
    \lambda_0^{(n,q)} &= \log \frac{\Pr( L_q^{(n)} = 0 | y_n, \hat{l}_{1:q-1}^{(n)})}{\Pr( L_q^{(n)} = 1 | y_n, \hat{l}_{1:q-1}^{(n)})}\\
    &= \log \frac{\Pr( L_q^{(n)} = 0, \hat{l}_{1:q-1}^{(n)} | y_n)}{\Pr( L_q^{(n)} = 1, \hat{l}_{1:q-1}^{(n)} | y_n)}.
\end{split}
\end{equation}
Let $\mathcal{L}_{1:q}(\hat{l}_{1:q-1}^{(n)},l)$ denote the assigned labels with the first $1$ to $q$ bits equal to $(\hat{l}_{1:q-1}^{(n)},l)$. We have
\begin{equation}
    \begin{split}
        \Pr( L_q^{(n)} &= l, \hat{l}_{1:q-1}^{(n)} | y_n)\\
        &= \sum_{R_{t} \in label^{-1}(\mathcal{L}_{1:q}(\hat{l}_{1:q-1}^{(n)},l))} \mathrm{Pr}(X_n\in R_{t}|y_n),
    \end{split}
\end{equation}
\begin{equation}
    \begin{split}
        & \mathrm{Pr}(X\in R_{t}|y) = \int_{r_{t-1}}^{r_{t}}\frac{1}{\sqrt{2\pi}\sigma_{X|y}}e^{-\frac{(x-\mu_{X|y})^{2}}{2\sigma^{2}_{X|y}}}dx\\
        & \qquad = \Phi \left( \frac{r_t - \mu_{X|y}}{\sigma_{X|y}} \right) - \Phi \left( \frac{r_{t-1} - \mu_{X|y}}{\sigma_{X|y}} \right),
    \end{split}
\end{equation}
where $\Phi(\cdot)$ is the cumulative distribution function of the standard normal distribution. Besides, $\mu_{X|y}$ and $\sigma_{X|y}$ denote the conditional mean value and the standard deviation of X given $Y=y$, which can be given as
\begin{equation}
    \begin{split}
        \mu_{X|y} = \frac{\sigma_Z^2}{\sigma_Z^2 + \sigma_W^2} y, \quad \sigma_{X|y}=\sqrt{\frac{2\sigma_{Z}^{2}\sigma_{W}^{2}+\sigma_{W}^{4}}{\sigma_{Z}^{2}+\sigma_{W}^{2}}}
    \end{split}
\end{equation}

Then Bob feeds $\lambda_0^{(n,q)}$ and $f_{1:M_q}^{(q)}$ to the SCL decoder to estimate $\hat{v}_{0:2N-1}^{(q)}$. Note that the permutation is necessary, as without it $u_n^{(q)}$ cannot be deconvolved even if $v_n^{(q)}$ is transmitted, since we do not have all the required preceding bits $u_{n-d:n-1}^{(q)}$.

\subsubsection{Key Extraction}
After the $q$-th decoding stage, Bob extracts the key $\hat{s}_{1:K_q}^{(q)}$ from the estimates $\hat{v}_{0:2N-1}^{(q)}$:
\begin{equation}
    \hat{s}_{1:K_q}^{(q)} = \hat{v}_{0:2N-1}^{(q)}(\mathcal{I}_{S}^{q}).
\end{equation}
The detailed code construction scheme will be elaborated in Section IV.

\section{Code Construction}
In this section, we propose a code construction scheme that considers both the polarization effect and the ML decoding performance. Although the proposed code construction algorithm incurs higher computational complexity, it can be precomputed offline. Importantly, adaptation to changing source statistics requires only rate selection without algorithm re-execution.

\subsection{Polarization Effect}
In order to utilize the polarization effect, density evolution (DE) is needed to evaluate the reliability of the equivalent synthesized channel $W(U_n^{(q)}|Y_{1:2N}, U_{0:2N-1}^{(1:q-1)}, U_{0:n-1}^{(q)})$ for every $q \in \{1, \dots, Q\}$. However, GA cannot be used directly because the virtual channel $W(L_q|Y,L_{q-1})$ is neither a binary erasure channel (BEC) nor a BI-AWGN channel. Fortunately, this can be solved by taking $W(L_q|Y,L_{q-1})$ as the BI-AWGN channel with the same capacity, i.e.,
\begin{equation}
    \begin{split}
        C\left(W\left(L_q|Y,L_{q-1}\right)\right) &= I(L_q; Y | L_{1:q-1})\\
        &= h(Y | L_{1:q-1}) - h(Y | L_{1:q}).
    \end{split}
\end{equation}
The entropies in (18) can be given as 
\begin{equation}
    \begin{split}
        & h(Y | L_{1:q}) = \sum_{l_{1:q}} \Pr(l_{1:q}) h(Y | L_{1:q} = l_{1:q}) \\
        &\approx \frac{1}{2^q} \sum_{y \in \mathcal{Y}} \sum_{l_{1:q}} -p(y | l_{1:q}) \log p(y | l_{1:q}) \Delta y\\
        &= \sum_{y \in \mathcal{Y}} \sum_{l_{1:q}} -p(y) \Pr(l_{1:q}|y) \left(q +\log (p(y) \Pr(l_{1:q}|y))\right) \Delta y,
    \end{split}
\end{equation}
where $\mathcal{Y}$ is the discretized set of the continuous variable $Y$ with distance $\Delta y$. $\Pr(l_{1:q}|y)$ can be computed via (13). Then the noise variance $\sigma_q^2$ of the equivalent BI-AWGN channel with the same capacity as the level-$q$ virtual channel $W(L_q|Y,L_{q-1})$ can be obtained according to the relationship as follows
\begin{equation}
    \begin{split}
        &C\left(W\left(L_q|Y,L_{q-1}\right)\right) = \\
        &\quad - \int_{-\infty}^{\infty} p_Y(y) \log_2 p_Y(y) \, dy - \frac{1}{2} \log_2(2 \pi e \sigma_q^2),
    \end{split}
\end{equation}
where $p_Y(y) = \frac{1}{2\sqrt{2\pi\sigma_q^2}} (e^{-\frac{(y-1)^2}{2\sigma_q^2}}+ e^{-\frac{(y+1)^2}{2\sigma_q^2}})$.

Assuming an all-zero $U_{0:2N-1}^{(q)}$, the bits LLRs $\Lambda_{m}^{(n,q)}, m = 1, \dots, \log2N, n = 0, \dots, 2^m-1$ in layer $m$ are approximated as Gaussian random variables with mean $\mathbb{E}(\Lambda_{m}^{(n,q)})$ and variance $\mathbb{V}(\Lambda_{m}^{(n,q)})$ satisfying $\mathbb{V}(\Lambda_{m}^{(n,q)}) = 2\mathbb{E}(\Lambda_{m}^{n,q})$. The recursive formulation of the mutual information between the LLR and the corresponding bits is \cite{Chiupolarml}
\begin{equation}
\begin{split}
    I_{m}^{(2i,q)} = 2I_{m-1}^{(i,q)} - J\left(\sqrt{2} J^{-1}(I_{m-1}^{(i,q)})\right),
\end{split}
\end{equation}
\begin{equation}
    I^{(2i+1,q)}_m = J\left(\sqrt{2} I_{m-1}^{(i,q)}\right),
\end{equation}
where $i=0, \dots, 2^{m-1}-1$ and
\begin{equation}
    J(\sigma) = 1 - \int_{-\infty}^{\infty} \frac{e^{-(x - \sigma^2 / 2)^2 / (2 \sigma^2)}}{\sqrt{2 \pi \sigma}} \log_2 (1 + e^{-x})dx.
\end{equation}
The mutual information for the initial layer is $I_{0}^{(0, q)} = J(2\rho)$, where $\rho$ is the signal-to-noise ratio (SNR) of the BI-AWGN channel. Then the reliability of the equivalent synthesized channel $W(U_n^{(q)}|Y_{1:2N}, U_{0:2N-1}^{(1:q-1)}, U_{0:n-1}^{(q)})$ can be evaluated via their mutual information $I_{\log2N}^{(0,q)}, \dots, I_{\log2N}^{(2N-1,q)}$.

\subsection{ML Decoding Performance}
The ML decoding performance is determined by the minimum Hamming distance of codewords and the multiplicity of this minimum distance. The source coding problem here can be interpreted as a channel coding problem by reversing the roles of ${L}$ and ${V}$. That is, ${L}$ can be taken as the output codeword of an input ${V}$ after deconvolution, inverse permutation, and polarization. The recursive relationship for deconvolution $u_{0:2N-1}^{(q)}=v_{0:2N-1}^{(q)}\mathbf{C}^{-1}$ is formulated as
\begin{equation}
    u_n = v_n \oplus \bigoplus_{m=1}^{\min(d,n)} c_m u_{n-m} \quad (n = 0, \cdots, 2N-1)
    \label{eq:placeholder}.
\end{equation}
$\mathbf{C}^{-1}$ is an upper triangle Toeplitz matrix given as
\begin{equation}
    \mathbf{C}^{-1} = 
    \begin{bmatrix}
        c_{0}' & c_{1}' & \cdots & c_{2N-1}' \\
        0 & c_{0}' & \cdots & c_{2N-2}' \\
        \vdots & \vdots & \ddots & \vdots \\
        0 & 0 & \cdots & c_{0}'
    \end{bmatrix}.
    \label{eq:matrix}
\end{equation}
The elements in $\mathbf{C}^{-1}$ can be derived in a recursive manner, i.e.,
\begin{equation}
    c_0' = 1, c_n' = \bigoplus_{m=1}^{\min(d,n)} c_m c_{n-m}' (n = 1, \cdots, 2N-1).
    \label{eq:dk}
\end{equation}

Then the inverse permutation, together with the polarization, can be formulated as $l_q^{(0:2N-1)} = u_{0:2N-1}^{(q)}\mathbf{G}_p^{-1}$. Treating $l_q^{(0:2N-1)}$ as the codeword generated by $v_{0:2N-1}^{(q)}$, the minimum distance of $L_q^{(0:2N-1)}$ is equal to the minimum weight of the non-zero codewords generated by $V_{0:2N-1}^{(q)}$ with $V_{0:2N-1}^{(q)}(\mathcal{I}_{F}^{q})=\bf{0}$. The upper bound approximation of the block error rate (BLER) under ML decoding reads
\begin{equation}
    {\rm{BLER}_{ML}} \approx W_{\min} Q\left( \sqrt{w_{\text{min}} \rho} \right)
\end{equation}
where $w_{\min}$ is the minimum weight of the non-zero codewords generated by $V_{0:2N-1}^{(q)}$ with $V_{0:2N-1}^{(q)}(\mathcal{I}_{F}^{q})=\bf{0}$. $W_{\text{min}}$ is the number of codewords with weight $w_{\min}$.

However, determining $w_{\min}$ and $W_{\min}$ is nontrivial and the traditional exhaustive search method suffers from prohibitive complexity. To address this, we propose a fast enumeration algorithm for efficient determination of $w_{\min}$ and $W_{\min}$, which is elaborated in the subsequence.

Let $\mathbf{g}_i^p$ represent the $i$th row of $\mathbf{G}_p^{-1}$. All nonzero $L_q^{(0:2N-1)}$ generated by $V_{0:2N-1}^{(q)}$ with $V_{0:2N-1}^{(q)}(\mathcal{I}_{F}^{q})=\bf{0}$ are combinations of elements in subsets of $\{\mathbf{g}_i^p, i \in [0,2N-1]\}$. All the possible combinations belong to $|\mathcal{I}_{S}^{q}|$ disjoint co-sets. The co-set led by $\mathbf{g}_i^p$ is
\begin{equation}
    \mathcal{C}_i \triangleq \left\{ \mathbf{g}_i^p \oplus \bigoplus_{j \in \mathcal{J}_i} \mathbf{g}_j^p : \mathcal{J}_i \subseteq [i + 1, 2N - 1] \right\}.
\end{equation}

For $i \in [0, 2N - 1]$ and $\mathcal{J}_i \subseteq [i + 1, 2N - 1]$, we have
\begin{equation}
    w(\mathbf{g}_i^p \oplus \bigoplus_{j \in \mathcal{J}_i} {\mathbf{g}_j^p}) = w({\mathbf{g}_{i'}} \oplus \bigoplus_{j \in \mathcal{J}_i} {\mathbf{g}_{j'}}) \ge w(\mathbf{g}_i^p) = w(\mathbf{g}_{i'}),
\end{equation}
where $\mathbf{g}_{i'}$ represents the $i'$th row of $\mathbf{G}^{\bigotimes \log{2N}}$ and $\mathbf{g}_i^p = \mathbf{g}_{i'}$ due to the permutation process. Define $\mathcal{S}_i$ as the support of the binary representation $\operatorname{bin}(i) = i_{\log{2N}-1} \ldots i_1 i_0$, that is, $\mathcal{S}_i \triangleq \{ \alpha \in [0, \log{2N} - 1] : i_\alpha = 1 \}$. We have ${\mathcal{S}_{i'}} = \{\log2N-1-s_i, s_i \in \mathcal{S}_i \}$.

Therefore, in order to count the minimum-weight $L_q^{(0:2N-1)}$, focusing only on the co-sets led by $\mathbf{g}_{i}^p$ with $i\in\{i \in \mathcal{I}_S^q: \mathbf{g}_{i}^p=w_{min}\}$ is enough. Furthermore, define
\begin{equation}
    \mathcal{K}_i^q = \{ j \in \mathcal{I}_S^q \setminus [0, i] \mid |\mathcal{S}_{j'} \setminus \mathcal{S}_{i'}| = 1 \},
\end{equation}
According to \cite{formation}, every $\mathcal{H}^q \subseteq \mathcal{K}_i^q$ together with additional rows forms a minimum-weight combination, which can be represented as
\begin{equation}
    w\left( \mathbf{g}_i^p \oplus \bigoplus_{h \in \mathcal{H}^q \cup \hat{\mathcal{H}}^{q}} \mathbf{g}_h^p \oplus \bigoplus_{m \in \mathcal{M}(\mathcal{H}^q \cup \hat{\mathcal{H}}^{q})} \mathbf{g}_m^p \right) = w_{\min},
\end{equation}
where $\hat{\mathcal{H}}^{q}$ is defined as
\begin{equation}
    \hat{\mathcal{H}}^{q} = \{ h \in \mathcal{I}_F^q \cap [i+1, 2N-1] : u_h = 1, |S_{h'} \setminus S_{i'}| = 1 \},
\end{equation}
which is induced by convolution. $\mathcal{K}_i^q$ and $\hat{\mathcal{H}}^{q}$ can be determined efficiently with the assistance of the addition and right-swap operations. Specifically, by the addition operation, all $j$ satisfying $w(\mathbf{g}_j^p) = 2w(\mathbf{g}_i^p)$ can be obtained by flipping each '0' in $\operatorname{bin}(i)$ one at a time. By the right-swap operation, all $j$ satisfying $w(\mathbf{g}_j^p) = w(\mathbf{g}_i^p)$ can be obtained by swapping each '1' in $\operatorname{bin}(i)$ with every '0' on the right one at a time. $\mathcal{M}(\mathcal{H}^{q} \cup \hat{\mathcal{H}}^{q})$ can be determined by the $\mathcal{M}$-construction as introduced in \cite{formation}.

The total number of minimum-weight combinations as described in (30) is equivalent to the number of subsets $\mathcal{H}^{q} \subseteq \mathcal{K}_{i}^{q}$, which is calculated as $\sum_{j=0}^{|\mathcal{K}_{i}^{q}|} \binom{|\mathcal{K}_{i}^{q}|}{j} = 2^{|\mathcal{K}_{i}^{q}|}$. However, not all these combinations with minimum weight $w_{min}$ exist. In addition to $\hat{\mathcal{H}}^{q}$, the rest set consisting of $a \in \mathcal{I}_F^q \cap [i+1, 2N-1]$ with $u_a=1$ induced by the convolutional structure can be defined as
\begin{equation}
    \mathcal{A} = \{a \in \mathcal{I}_F^q \cap [i+1, 2N-1]: u_a=1, \mathcal{S}_{a'} \setminus \mathcal{S}_{i'} > 1\}.
\end{equation}
We discount all $\mathcal{H}^{q} \subseteq \mathcal{K}_{i}^{q}$ for which there exist $a \in \mathcal{A}$ but $a \notin \mathcal{M}(\mathcal{H}^{q} \cup \hat{\mathcal{H}}^{q})$, or $m \in \mathcal{M}(\mathcal{H}^{q} \cup \hat{\mathcal{H}}^{q}) \cap \mathcal{I}_F^q$ but $m \notin \mathcal{A}$, that is, the minimum-weight combinations in (30) do not exist. To achieve this efficiently, we define $\mathcal{B} = \{b \in \mathcal{I}_F^q \cap [i+1, 2N-1]: \mathcal{S}_{b'} \setminus \mathcal{S}_{i'} > 1\}$ and $b_{\max} = \max\{\mathcal{B}\}$. Then $\mathcal{K}_{i}^{q}$ can be partitioned into two sets: $\hat{\mathcal{K}}_{i}^q = \{ j \in \mathcal{K}_{i}^{q} : j < b_{\text{max}} \}$ and its complementary set. Since the minimum-weight combination exists for every $\mathcal{H}^{q} \subseteq \mathcal{K}_{i}^{q} \setminus \hat{\mathcal{K}}_i^q$, we have $2^{|\mathcal{K}_{i}^{q} \setminus \hat{\mathcal{K}}_i^q|}$ minimum-weight codewords for subsets $\mathcal{H}^{q} \subseteq \mathcal{K}_{i}^{q} \setminus \hat{\mathcal{K}}_i^q$. The problem is thus reduced to enumerating and discounting $\mathcal{H}^{q} \subseteq \hat{\mathcal{K}}_i^q$ from the $2^{|\hat{\mathcal{K}}_i^q|}$ subsets of $\hat{\mathcal{K}}_i^q$ for which the minimum-weight combinations do not exist. Finally, the number of minimum-weight combinations that exist within $\mathcal{C}_i$ can be obtained as
\begin{equation}
    W_{i,\min} = 2^{|\mathcal{K}_{i}^{q} \setminus \hat{\mathcal{K}}_i^q|} (2^{|\hat{\mathcal{K}}_i^q|} - \hat{W}_i),
\end{equation}
where $\hat{W}_i$ is the number of subsets $\mathcal{H}^{q} \subseteq \hat{\mathcal{K}}_i^q$ for which the minimum-weight combinations do not exist.

\subsection{Code Construction Algorithm}
\begin{figure}[!t]
    \begin{algorithm}[H]
\caption{Code construction algorithm}
\label{alg:polar-selection}
\begin{algorithmic}[1]
\STATE \textbf{Input} $\text{BLER}_{T}$, $\beta$, and $N$
\STATE \textbf{Initialization} \\
$e_{0}^* \gets 2N-1$, $\mathcal{E}_0^* \gets \{e_0\}$,\\
$\rho_{0}^* \gets \dfrac{1}{2N}\left[Q^{-1}\left(\text{BLER}_{T}\right)\right]^{2}$
\FOR{$t = 1$ \textbf{to} $2N-1$}
    \STATE Update by (20) and (21) at SNR $\rho_{t-1}$ to obtain:\\
    $I_{\log2N,\rho_{t-1}}^{(0)},I_{\log2N,\rho_{t-1}}^{(1)},\ldots,I_{\log2N,\rho_{t-1}}^{(2N-1)}$
    \STATE Select $\mathcal{Z} \subseteq \{0,1,\ldots,2N-1\} \setminus \mathcal{E}_{t-1}^*$ with $|\mathcal{Z}| = \min(\beta,2N-t)$ such that $I_{\log2N,\rho_{t-1}}^{(z)} \geq I_{\log2N,\rho_{t-1}}^{(\hat{z})}$, $\forall z \in \mathcal{Z}$ and $ \forall \hat{z} \in \{0,1,\ldots,2N-1\} \setminus\{\mathcal{E}_{t-1}^*\cup\mathcal{Z}\}$
    \FOR{all $z \in \mathcal{Z}$}
    \STATE Set $e_{t} \gets z$, $\mathcal{E}_{t} \gets \{\mathcal{E}_{t-1}^*,e_{t}\}$, $\rho_t^* \gets 0$
    \FOR{all $i \in \{i \in \mathcal{E}_{t}: \mathbf{g}_i^p = w_{\min} = \underset{i\in \mathcal{E}^{(z)}}{\min}2^{|\mathcal{S}_{i'}|}\}$}
    \STATE Determine $W_{i,\min}$ with reference to Algorithm 1 in \cite{fastenum}
    \STATE $W_{\min} \gets W_{\min} + W_{i,\min}$
    \ENDFOR
    \STATE Set $\rho_{t} \gets \dfrac{1}{w_{\min}} \left[ Q^{-1}\left( \dfrac{\text{BLER}_{T}}{W_{\min}} \right) \right]^{2}$
    \IF{$\rho_t \le \rho_t^*$}
    \STATE Update $\rho_t^* \gets \rho_t$, $e_{t}^* \gets z$, $\mathcal{E}_{t}^* \gets \mathcal{E}_{t}$
    \ENDIF
    \ENDFOR
\ENDFOR
\STATE \textbf{Output} $e_{0:2N-1}^* \gets [e_{0}^*, e_{1}^*, \ldots, e_{2N-1}^*]$
\end{algorithmic}
\end{algorithm}
\end{figure}

Considering both the polarization effect and the ML decoding performance, a greedy algorithm is developed to design the source PAC codes for all code rates $K_q/2N$, $K_q \in \{1, 2, \ldots, 2N\}$ by constructing an ordered sequence of bit-channel indices $e_{0:2N-1}^* \triangleq [e_{0}^*, e_{1}^*, \ldots, e_{2N-1}^*]$. The specific process is summarized in Algorithm 1. In each step, a new index is appended to $e$. In the initial step, we optimally set $e_{0}^*=2N-1$ and obtain the required SNR $\rho_0^*$ to achieve the preset BLER. In the $t$th step, the mutual information $I_{\log2N,\rho_{t-1}}^{(0:2N-1)}$ is updated with the required SNR $\rho_{t-1}$. Then we select a set $\mathcal{Z} \subseteq \{0,1,\ldots,2N-1\} \setminus \mathcal{E}_{t-1}^*$ consisting of $|\mathcal{Z}| = \min(\beta,2N-t)$ indices with the highest values of mutual information, where $\mathcal{E}_{t-1}^*$ is defined as the set of the chosen indices after step $t-1$. For each $z \in \mathcal{Z}$, we set $e_{t} = z$ and $\mathcal{E}_t = \{\mathcal{E}_{t-1}^*,e_{t}\}$. Then the fast enumeration is performed to evaluate the required SNR $\rho_t$ with the unfrozen set $\mathcal{E}_t$. We choose the optimal $z$ and set $\rho_t^* = \rho_t$, $e_{t}^* = z$, and $\mathcal{E}_t^* = \mathcal{E}_t$. 

After obtaining the ordered sequence $e_{0:2N-1}^*$ of length $2N-1$, we can adaptively select $K_q$ according to the noise variance $\sigma_q^2$ of the equivalent BI-AWGN channel and the reliability constraint. Then the unfrozen set is determined as $\mathcal{I}_S^q = \mathcal{E}_{K_q-1}^*$. The control parameter $\beta$ governs the balance between the exploitation of polarization and ML decoding performance.

\section{SIMULATION RESULTS}
In this section, simulations are carried out to evaluate the performance of the proposed scheme. The SNR in this section is defined as $\sigma_h^2/\sigma_n^2$. The quantization level is set as $Q = 8$ and the list size is set as 64.

\begin{figure}[!t]
\centerline{\includegraphics[width=0.9\linewidth]{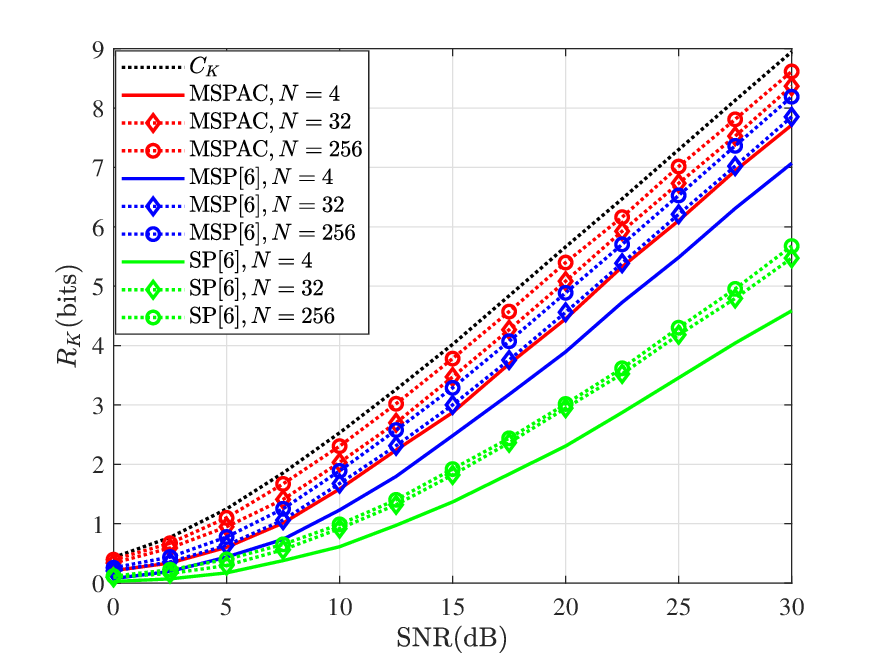}}
\caption{Comparison of different scheme.}
\label{fig1}
\end{figure}

Fig. 1 illustrates the key generation rate $R_K$ versus the SNR under different schemes with $N = \{4, 32, 256\}$. $\text{BDR} \le \epsilon_{\text{BDR}}$ with $\epsilon_{\text{BDR}}=10^{-2}$ is set as the reliability constraint and the parameter $\beta$ is set to 5. The proposed multilevel source PAC coding (MSPAC) is shown to significantly outperform the multilevel source polar coding (MSP) in \cite{shortpolar} and the source polar coding (SP) in terms of key generation rate. There are two reasons for this superiority. First, the outer convolutional transform compensates for the capacity loss induced by the incomplete polarization at short blocklength. Second, the hierarchical structure of multilevel source coding inherently enhances the polarization effect and provides an additional polarization gain, which outweighs the loss induced by a shorter blocklength at each level.

\begin{figure}[!t]
\centerline{\includegraphics[width=0.9\linewidth]{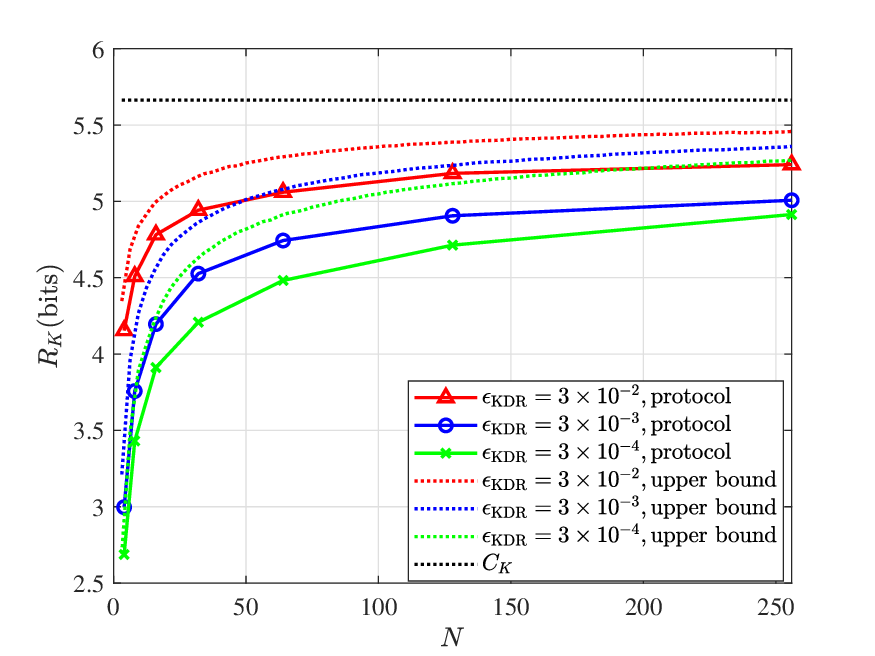}}
\caption{Comparison of $R_K$ achieved by the scheme and the bound.}
\label{fig2}
\end{figure}

Fig. 2 demonstrates the comparison between the key generation rate $R_K$ achieved by the scheme and the bound versus $N$. The reliability constraint is set as $\text{KDR} \le \epsilon_{\text{KDR}}$ with $\epsilon_{\text{BDR}}=\{3\times10^{-2},3\times10^{-3},3\times10^{-4}\}$. The SNR is set to 20 dB and the upper bound is obtained conservatively by setting $\delta = 0$. It can be seen that the gap between the key generation rate achieved by the scheme and the bound is relatively small. The key generation rate $R_K$ increases with the decrease of $\epsilon_{\text{BDR}}$. As $N$ increases, $R_K$ increases and demonstrates progressively diminishing gain. In addition, the upper bound is shown to be tighter than $C_K$.

\balance

\section{Conclusion}
This paper developed a key generation scheme based on multilevel source PAC codes. In order to adapt to the low-latency constraints of key generation in time-sensitive applications, the keys were generated from limited continuous source observations, and the source coding performance at short blocklength was investigated. A novel joint-polarization-distance code construction algorithm was also proposed. The simulation results demonstrated that the proposed scheme enables a superior key rate closer to the upper bound and indicated its applicability for low-latency key generation.

\section{Acknowledgement}
This study was supported by the National Science Foundation of China (NSFC) under grant U22A2001, the Henan Natural Science Foundation for Excellent Young Scholar under Grant 242300421169, and the Fundamental Research Funds for the Central Universities, No. B240203012.
\bibliographystyle{IEEEtran}
\bibliography{reference}

\begin{thebibliography}{10}
\providecommand{\url}[1]{#1}
\csname url@samestyle\endcsname
\providecommand{\newblock}{\relax}
\providecommand{\bibinfo}[2]{#2}
\providecommand{\BIBentrySTDinterwordspacing}{\spaceskip=0pt\relax}
\providecommand{\BIBentryALTinterwordstretchfactor}{4}
\providecommand{\BIBentryALTinterwordspacing}{\spaceskip=\fontdimen2\font plus
\BIBentryALTinterwordstretchfactor\fontdimen3\font minus
  \fontdimen4\font\relax}
\providecommand{\BIBforeignlanguage}[2]{{%
\expandafter\ifx\csname l@#1\endcsname\relax
\typeout{** WARNING: IEEEtran.bst: No hyphenation pattern has been}%
\typeout{** loaded for the language `#1'. Using the pattern for}%
\typeout{** the default language instead.}%
\else
\language=\csname l@#1\endcsname
\fi
#2}}
\providecommand{\BIBdecl}{\relax}
\BIBdecl

\bibitem{one-shot}
C.~T. Li and V.~Anantharam, ``One-shot variable-length secret key agreement
  approaching mutual information,'' \emph{IEEE Transactions on Information
  Theory}, vol.~67, no.~8, pp. 5509--5525, Aug. 2021.

\bibitem{6G}
R.~Liu, L.~Zhang, R.~Y.-N. Li, and M.~D. Renzo, ``The itu vision and framework
  for 6g: Scenarios, capabilities, and enablers,'' \emph{IEEE Vehicular
  Technology Magazine}, vol.~20, no.~2, pp. 114--122, Jun. 2025.

\bibitem{lattices}
L.~Luzzi, C.~Ling, and M.~R. Bloch, ``Optimal rate-limited secret key
  generation from gaussian sources using lattices,'' \emph{IEEE Transactions on
  Information Theory}, vol.~69, no.~8, pp. 4944--4960, Aug. 2023.

\bibitem{commcontrainedkey}
H.~Hentilä, Y.~Y. Shkel, and V.~Koivunen, ``Communication-constrained secret
  key generation: Second-order bounds,'' \emph{IEEE Transactions on Information
  Theory}, vol.~70, no.~11, pp. 8180--8203, Nov. 2024.

\bibitem{chousourcepolar}
R.~A. Chou, M.~R. Bloch, and E.~Abbe, ``Polar coding for secret-key
  generation,'' \emph{IEEE Transactions on Information Theory}, vol.~61,
  no.~11, pp. 6213--6237, Nov. 2015.

\bibitem{shortpolar}
H.~Hentilä, Y.~Y. Shkel, and V.~Koivunen, ``Secret key generation using short
  blocklength polar coding over wireless channels,'' \emph{IEEE Journal of
  Selected Topics in Signal Processing}, vol.~16, no.~1, pp. 144--157, Jan.
  2022.

\bibitem{rowmerged}
A.~Zunker, M.~Geiselhart, L.~Johannsen, C.~Kestel, S.~ten Brink, T.~Vogt, and
  N.~Wehn, ``Row-merged polar codes: Analysis, design, and decoder
  implementation,'' \emph{IEEE Transactions on Communications}, vol.~73, no.~1,
  pp. 39--53, Jan. 2025.

\bibitem{polarrule1}
X.~Yao, X.~Zheng, and X.~Ma, ``Trellis-based construction of polar codes for
  scl decoding,'' in \emph{2024 IEEE International Symposium on Information
  Theory (ISIT)}, 2024, pp. 2210--2215.

\bibitem{rmpolar}
M.~Rowshan, A.~Burg, and E.~Viterbo, ``Polarization-adjusted convolutional
  (pac) codes: Sequential decoding vs list decoding,'' \emph{IEEE Transactions
  on Vehicular Technology}, vol.~70, no.~2, pp. 1434--1447, Feb. 2021.

\bibitem{Chiupolarml}
M.-C. Chiu and Y.-S. Su, ``Design of polar codes and pac codes for scl
  decoding,'' \emph{IEEE Transactions on Communications}, vol.~71, no.~5, pp.
  2587--2601, May 2023.

\bibitem{formation}
M.~Rowshan, S.~H. Dau, and E.~Viterbo, ``On the formation of min-weight
  codewords of polar/pac codes and its applications,'' \emph{IEEE Transactions
  on Information Theory}, vol.~69, no.~12, pp. 7627--7649, Dec 2023.

\bibitem{fastenum}
M.~Rowshan and J.~Yuan, ``Fast enumeration of minimum weight codewords of pac
  codes,'' in \emph{2022 IEEE Information Theory Workshop (ITW)}, 2022, pp.
  255--260.

\end{thebibliography}

\end{document}